\documentclass[11pt]{article}
\usepackage{amsmath}
\usepackage{amsfonts}
\usepackage{amssymb}
\usepackage{hyperref}
\usepackage{latexsym}
\usepackage[dvips]{graphicx}
\usepackage{epsf}
\usepackage{bbm}

\newcommand{\be}{\begin{equation}}
\newcommand{\ee}{\end{equation}}
\newcommand{\bea}{\begin{eqnarray}}
\newcommand{\eea}{\end{eqnarray}}
\newcommand{\nn}{\nonumber}

\let\a=\alpha \let\b=\beta  \let\g=\gamma  \let\d=\delta
   
\let\m=\mu    \let\n=\nu          \let\r=\rho \let\om=\omega
\let\s=\sigma \let\vphi=\varphi \let\Si=\Sigma 
 \let\eps=\epsilon

\newcommand{\f}{\frac}

\def\tl{\tilde}
\newcommand{\Id}{{\mathbbm 1}}
\newcommand{\p}{\partial}
\newcommand{\na}{\nabla}

\newcommand{\w}{\wedge}

\newcommand{\Ref}[1]{(\ref{#1})}

\newcommand{\scr}{\rm\scriptscriptstyle}
\def\Pp{P_{\scr (+)}}\def\Pm{P_{\scr (-)}}

\newcommand{\wh}{\hat{w}}
\newcommand{\ws}{\check{w}}
\newcommand{\wb}{\bar{w}}
\newcommand{\Ph}{\hat{P}}
\newcommand{\Ps}{\check{P}}
\newcommand{\Pb}{\bar{P}}

\def\C{{\mathbbm C}}

\begin{document}

\title{\bf Pauli-Fierz mass term in modified Plebanski gravity}
\author{{David Beke$^{a}$, Giovanni Palmisano$^{b}$ and Simone Speziale$^{c}$}
\smallskip \\ 
{\small $^a$ \emph{Ghent University, Department of Mathematical Analysis EA16, Galglaan 2, 9000 Ghent, Belgium}}\\
{\small $^b$ \emph{Dipartimento di Fisica dell'Universit\`a ``La Sapienza", INFN Sez.\,Roma1, I-00185 Roma, Italy}} \\
{\small $^c$\emph{Centre de Physique Th\'eorique, Aix-Marseille University, CNRS-UMR 7332, 13288 Marseille, France}}
}

\maketitle

\begin{abstract}
We study SO(4) BF theory plus a general quadratic potential, which describes a bi-metric theory of gravity.
We identify the profile of the potential leading to a Pauli-Fierz mass term for the massive graviton, thereby avoiding the linearized ghost. We include the Immirzi parameter in our analysis, and find that the mass of the second graviton depends on it. 
At the non-perturbative level, we find a situation similar to genuine bi-gravities: even choosing the Pauli-Fierz mass term, the ghost mode propagates through the interactions. We present some simple potentials leading to two and three degrees of freedom, and discuss the difficulties of finding a ghost-free bi-gravity with seven degrees of freedom.
Finally, we discuss alternative reality conditions for the case of SO(3,1) BF theory, relevant for Lorentzian signature, and give a new solution to the compatibility equation.
\end{abstract}

\section{Introduction}

Topological theories of the BF type bear interesting relations with general relativity, see \cite{BFreview} for a recent review. 
A mechanism to relate the two theories is the use of the Plebanski action, which comes in either the self-dual \cite{Plebanski, Capo2} or the non-chiral \cite{Capo2,Mike,depietri} formalism. In the latter, which is the subject of this paper, one starts with BF theory with the Lorentz group (or SO(4) in Euclidean signature) as local gauge group, and adds suitable ``simplicity'' constraints to recover the dynamics of general relativity. 
  The resulting action is polynomial, and it plays an important role in the spin foam approach to quantum gravity \cite{Ashtekar,Baez,Perez,reviews}.
	
Recently \cite{Speziale}, building on previous work in the self-dual formalism \cite{Krasnov1,Freidel,KrasnovEff}, it was shown that SO(4) BF theory -- alone, without constraints -- can be parametrized in terms of two metrics plus scalars fields. 
The metric interpretation is however completely artificial, and carries no physics, because of the large ``shift'' symmetry that makes the theory topological: any pair of metric is connected to the bi-flat one (e.g. for trivial topology) by a shift transformation, and no local degrees of freedom propagate.
The situation changes when the simplicity constraints are added, and general relativity is recovered. The advantage of the bi-metric parametrization is to make the role of the constraints completely transparent:
they freeze the scalars fields, and identify the two metrics with one another. 

An interesting set of theories can be constructed by adding a potential term to the BF action, instead of the simplicity constraints. 
Such modified non-chiral Plebanski actions were proposed initially in \cite{Lee} as models for grand unification, and in \cite{Krasnov1,Krasnov} for the self-dual case.\footnote{In the self-dual case, the modification turns out to be related to a class of generally covariant actions previously considered in \cite{Capo3,BengtssonMod}. The modified self-dual theory has been studied quite extensively in the literature, e.g. \cite{Krasnov1,KrasnovEff,Krasnov2,Krasnov3,Ishibashi}.}
The addition of the potential term breaks the shift symmetry, and propagating local degrees of freedom (DOFs) appear. A canonical analysis performed in \cite{Alexandrov} showed that for a generic potential, there are eight DOFs. Their interpretation was clarified in \cite{Speziale}, where it was shown that the action describes a bi-metric theory of gravity plus scalars, with the two metrics interacting through the scalars and the potential. The eight degrees of freedom turned out to be, at the linearized level, the same of standard bi-metric theories \cite{Isham,Damour}. Namely, a massless and a massive graviton, plus a massive scalar. 
The scalar mode found in \cite{Speziale} is a ghost, and the theory is unstable around the bi-flat solution. This is a situation familiar from massive gravity, where a ghost is present unless the particular Pauli-Fierz form \cite{Fierz} for the mass term is chosen. 
The main result of the present paper is to complete the analysis of \cite{Speziale}: we study the most general quadratic potential of the modified Plebanski action, and identify the values of the coefficients leading to the Pauli-Fierz mass term. 
In doing so, we also extend the previous analysis to include the Immirzi parameter, and find that the mass of the second graviton depends on it. 
This might be surprising at first sight, but it is a natural consequence of giving a canonical form to the kinetic terms of the action, and it suggests an intriguing new role for the Immirzi parameter in such modified theories of gravity.
We also give a more complete description of the linearized theory, studying the decomposition of the various fields in their irreducible representations, and unravel a simple relation between SO(4) BF theory and general relativity at this level: loosely speaking, second order SO(4) BF theory can be written as Einstein's $\Gamma\Gamma$ action but with the connection as a functional of two metrics and scalar fields. This material appears in sections 2 to 4.

The ghost absence at the linearized level is however not enough to guarantee the stability of the theory, as the unhealthy degree of freedom can reappear through interactions, as it is well-known from a classic result by Boulware and Deser \cite{Boulware}. While it has been argued that the theory is stable on cosmological or Lorentz-violating backgrounds \cite{Damour,Blas,Nesti,Banados,Milgrom}, it is interesting to understand whether flat background stability is necessarily compromised by a graviton mass. In particular, recent literature \cite{deRham,Hassan1,Hassan2} claims that a specific, non-polynomial potential is non-perturbatively ghost-free, although a debate on the proof still continues \cite{HassanNew}. 
One might hope that the situations is different in the context of the modified Plebanski theory, due to the presence of additional scalar fields. However, this is not the case. In Section 5, we show that the Pauli-Fierz mass term is again not a sufficient condition for the non-perturbative absence of the ghost. 
Although we are not able to provide a systematic study of all possible potentials, we discuss a few natural choises.
While classes with two or three degrees of freedom can be easily identified, classes with seven degrees of freedom, corresponding to ghost-free bigravities, appear elusive. At least, we are able to point out a specific algebraic difficulty for this. Merely adapting the potential of  \cite{deRham,Hassan1,Hassan2} to the Plebanski formalism is possible, but it does not appear any less artificial, nor can it easily be used for an independent test of ghost-freeness.

Finally, the extension of these results to the Lorentzian signature is straightforward, but requires the use of complex fields, and suitable reality conditions. These are the same present also in the self-dual formalism, and are difficult to deal with at the quantum level. The non-chiral formulation has the advantage that one can use alternative metrics which are automatically real, without the need of auxiliary conditions. To illustrate the interest of this interpretation, in section 5 we show how one can use these metrics to give a new exact solution to the compatibility equations that does not require any reality conditions.

\section{Plebanski formalism and modified theories of gravity}

Our starting point is BF theory with local gauge group SO(4). Extension to the case SO(3,1), relevant for gravity in Lorentzian signature, will be discussed below in section \ref{secLor}.
The fundamental fields are a connection $\om$ in the local Lorentz gauge group, and an algebra-valued 2-form $B$. 
The action is
\be\label{SBF}
S_{BF}(B,\om) = \int P_\g^{IJKL} B_{IJ} \w F_{KL}(\om), \qquad P_\g^{IJKL} = \d^{IJKL} +\f1{2\g}\eps^{IJKL},
\ee
with $\g$ a real, dimensionsless parameter.
The theory is invariant under the shift symmetry $(\d_\alpha B=d_\om\alpha, \d_\alpha \om=0)$, where $\alpha$ is an algebra-valued 1-form. The field equations are
\bea
d_\om B =0\label{dB}, \qquad
F(\om)=0,
\eea
where we used the fact that the Hodge star $\star=\f12\eps^{IJ}_{KL}$ commutes with the covariant exterior derivative.
The equations do not allow for local degrees of freedom:  under a shift transformation, all solutions are locally equivalent,
e.g. to $B=\om=0$ for trivial topology.

BF theory can be related to (modified or not) theories of gravity considering the following action,
\be\label{S}
S(B,\om,\phi) = \int P_\g^{IJKL} B_{IJ} \w F_{KL}(\om) + \Big(\phi_{IJKL}-\f12\Lambda(\phi)\eps_{IJKL}\Big) B^{IJ}\w B^{KL},
\ee
where $\Lambda(\phi)$ does not contain derivatives.
The potential generically breaks the shift symmetry, leaving local gauge and diffeomorphisms as symmetries, and introduces non-trivial curvature: varying with respect to $B$ we now get
\be
F(\om) = -2 \Big(\phi-\Lambda(\phi) \, \star\Big) B,
\ee
instead of \Ref{dB}.
Crucial are the field equations coming from the variation with respect to the new field $\phi$.
This tensor is taken to be symmetric under exchange of the first and second pair of indices, and antisymmetric in each pair. 
It thus has 21 independent components, which can be classified in terms of irreducible representations of SO(4), as
\be\label{dec}
\phi \in [{\bf (1,0)}\oplus{\bf (0,1)}]\otimes_{\rm sym}[{\bf (1,0)}\oplus{\bf (0,1)}]
= {\bf (2,0)}\oplus{\bf (0,2)}\oplus{\bf (1,1)}\oplus{\bf (0,0)}\oplus{\bf (0,0)}.
\ee
The irreducibles are the ``Weyl'' components, that is the symmetric and traceless tensors \linebreak $(\vphi_{ij}, \bar\vphi_{ij})\in{\bf (2,0)}\oplus{\bf (0,2)}$, the trace-free ``Ricci'' components $\psi_{ij} \in {\bf (1,1)}$, and the two traces, $\phi=\d_{IJKL}\phi^{IJKL}$ and $\phi_\star=(1/2)\eps_{IJKL}\phi^{IJKL}$.
See appendix \ref{App} for the explicit decomposition. It is furthermore assumed that $\phi_\star=0$, as otherwise variation of (\ref{S}) with respect to $\phi$ would restrict us to the case $B^{IJ}\wedge B_{IJ}=B^{IJ}\wedge\star B_{IJ}=0$, which is degenerate  in a sense to be clarified later. The action \Ref{S} describes a general class of theories, of which general relativity in the Plebanski formalism is a special case, obtained for $\Lambda(\phi)=\Lambda$ constant. In this case, the variation with respect to $\phi$ imposes the simplicity constraints, and the action is equivalent to general relativity in the first order formalism, with Immirzi parameter $\g$ and cosmological constant proportional to $\Lambda$. See e.g. \cite{depietri, PlebImm} for details on the non-chiral Plebanski formulation.

The number of degrees of freedom of \Ref{S} has been studied canonically \cite{Alexandrov}.
When the Hessian matrix
\be\label{hessian}
\Lambda^{(2)}{}^{IJKL}_{MNPQ}=\f{\d^2 \Lambda(\phi)}{\d \phi_{IJKL} \d \phi^{MNPQ}}
\ee
has maximal rank, there are eight DOFs. Hence an invertible Hessian corresponds to a theory with six DOFs more than the two DOFs of Plebanski gravity.\footnote{For the reader familiar with the canonical analysis of the Plebanski action, this result can be understood as follows:
the variation of \Ref{S} with respect to $\phi^{IJKL}$ gives algebraic constraints on the components of $\phi$. When $\Lambda^{(2)}$ has zero rank, the case of a constant potential term, there are still algebraic constraints (on components of $B$ this time), but six of these equations now generate secondary constraints. The resulting system is second class, and six degrees of freedom are killed, reducing the DOFs to only two. For generic $\Lambda$, the lack of the six constraints is the source of the six extra degrees of freedom, for a total of 8.}
To obtain a theory with fewer degrees of freedom, special potentials with singular Hessians are needed. To complicate the analysis, the zero modes are required to be in a specific six-dimensional subspace \cite{Alexandrov}. While a systematic study of the singular Hessian has not appeared in the literature, below in Section \ref{SecNP} we present a survey of simple scenarios.

As a case study, we consider quadratic modifications, more precisely a potential of the form 
$$\Lambda(\phi)=\Lambda+(\phi, A\phi),$$ 
in matricial notation. 
While this is of course a strong limitation on all the possible potential terms, it turns out to be large enough for our purposes, for both perturbative and non-perturbative considerations.
Given the local SO(4) invariance, there are 
four quadratic invariants,\footnote{See e.g. \cite{Carminati}. In the appendix we provide the complete list of invariants without symmetry assumptions.}
\begin{subequations}\label{Qinv}\bea
&& Q_1(\phi) = \phi_{IJKL} \phi^{IJKL} 
= \vphi_{ij}^2 + \bar\vphi_{ij}^2 + \f12 \psi_{ij}^2 +\f16 \phi^2, \\ 
&& Q_2(\phi) = \f12 \eps^{KL}_{MN} \phi_{IJKL} \phi^{IJMN} 
= \vphi_{ij}^2 - \bar\vphi_{ij}^2, \\ 
&& Q_3(\phi) = \phi_{IK} \phi^{IK} 
= \f14 \phi^2 +\f14 \psi_{ij}^2. \\
&& Q_4(\phi) = \phi^2,
\eea\end{subequations}
where $\phi^{IK} = \d_{JL} \phi^{IJKL}$.
It is convenient to take the irreducible components as basis, and to parametrize the potential as
\be\label{fAf}
(\phi, A\phi) = \f1{a_1} \, \vphi_{ij}^2 + \f1{a_2} \, \bar\vphi_{ij}^2 + \f1{2 a_3} \, \psi_{ij}^2 + \f1{6 a_4} \, \phi^2.
\ee
We assume for the moment that all coefficients are non-vanishing.
Therefore, the Hessian is non-singular, and we expect 8 DOFs from \cite{Alexandrov}.
At the same time, the gradient
\be\label{gradL}
\f12 \frac{\d \Lambda}{\d \phi_{IJKL}}=\f1{a_1} \Pi_{\bf(2,0)}^{IJKL}{}_{ij} \varphi^{ij}+ \f1{a_2} \Pi_{\bf(0,2)}^{IJKL}{}_{ij} \bar{\varphi}^{ij} +\f1{2 a_3} \Pi_{\bf(1,1)}^{IJKL}{}_{ij} \psi^{ij}+ \f1{6 a_4} \d^{IJKL}\phi,
\ee
can be solved for all components of $\phi^{IJKL}$, hence it is possible to completely eliminate the field from the action, using its field equations.
Introducing the short-hand notation
\be\label{mdef}
Q^{IJKL} = B^{IJ}\w B^{KL},  
\ee
the variation of \Ref{S} with respect to $\phi$ gives
\be\label{varphi}
Q^{IJKL} - \f1{12} \eps^{IJKL} Q_\star = 2 Q_\star (A\phi)^{IJKL},
\ee
where $Q_\star = (1/2) \eps_{IJKL} Q^{IJKL}$. The scalar density $Q^{IJKL}$ decomposes as \Ref{dec}. Accordingly, we denote $Q_{ij}$, $\bar Q_{ij}$, ${\mathbbm Q}_{ij}$, $Q$ and $Q_\star$ the various irreducible representations. 
Assuming $Q_\star\neq 0$, \Ref{varphi} is easily inverted.
Inserting this solution back into the action yields a $\phi$-independent form,
\begin{subequations}\bea\label{Sgen}
S(B,\om) &=& \int P_\g^{IJKL} B_{IJ} \w F_{KL} + V(B),
\label{Sgen1} \\ \label{Vm}
V(B) &=& 
-\Lambda Q_\star + \f1{4Q_\star} \left( a_1 \, Q_{ij}^2 + a_2 \, \bar Q_{ij}^2 + \f12 a_3 \, {\mathbbm Q}_{ij}^2 + \f16 a_4 \, Q^2 \right).
\eea\end{subequations}

For general $\Lambda(\phi)$, as long as its Hessian is  non-singular, $\phi^{IJKL}$ can be eliminated to yield an action of the form \Ref{Sgen}, with $V(B)$ of the form
\be
V(B)=Q_\star \, V\Big( ({Q^{IJKL} - \f1{12} \eps^{IJKL} Q_\star})/{Q_\star}  \Big).
\ee
This restricted form is the one compatible with the symmetries of the field $\phi$ that has been integrated out, which also explains the absence of terms in $Q_\star^2$ and $QQ_\star$ in the numerator, and $Q$ in the denominator. Of course for singular Hessians -which in the quadratic case appear if any of the coefficients in \Ref{fAf} vanishes- this equivalence is broken.\footnote{It is also of interest to take actions of the form \Ref{Sgen1} as the starting point, with $V(B)$ for instance given by the most general homogeneous function of degree 1 of $Q$. A new canonical analysis is then necessary, but it is easy to see that a generic potential has again 8 DOFs, and a system with fewer needs the presence of zero modes in a certain Hessian matrix analogue to \Ref{hessian}. Our preference to work with \Ref{S} has two reasons: (i), the link with general relativity is more natural (the constant $\Lambda$ case versus a Dirac-delta potential), and (ii), having an additional field, $\phi$, at disposal, can only make the study of singular potentials easier allowing for more configurations.}
The case considered in \cite{Speziale}, $V(B) = \f{A}{2 Q_\star} (Q_{IJKL}^2 - \f16 Q_\star^2)$, corresponds to the single $Q_1(\phi)$ invariant.
It is a special case of the analysis in the present paper, that can be found by setting $a_i\equiv 1/2A \ \forall i$. 
For this case, perturbation theory around the bi-flat solution is unstable, which is what we want to cure in this paper.

\subsection{Bi-metric interpretation}

The metric enters the Plebanski formalism only as a derived quantity, through a cubic relation with the $B$ fields. In the same sense in which the tetrad can be called the ``square root'' of the metric, the $B$ field is its ``cubic root'' \cite{thooft}.
What is special about the SO(4) formalism, with respect to the original self-dual case, is that there are two different combinations of the $B$ field that define a metric, that is
\be\label{gpm}
\sqrt{g^{\scr (\pm)}} \, g^{\scr (\pm)}_{\mu\nu} = 
\f1{6} \d_{IN}\left(\d_{JMKL} \pm \f12 \eps_{JMKL}\right) B^{IJ}_{\mu\a}\tl B^{KL \a\b}B^{MN}_{\b\nu},
\ee
where $\tilde B^{\a\b}=\f12\eps^{\a\b\g\d} B_{\g\d}$. These correspond to the Urbantke metrics \cite{Capo2,Urbantke} of the left and right $\mathfrak{su(2)}$ subalgebras.

A strategy to parametrize the $B$ field in terms of these two metrics is as follows: first, split the algebra, $\mathfrak{so(4)=su(2)\oplus su(2)}$, using projectors $P^{IJ}_{\scr (\pm)}{}_i$ (See appendix \ref{App} for conventions and full details), thus
$B^{IJ}=P^{IJ}_{\scr (+)}{}_i B^i_{\scr (+)} + P^{IJ}_{\scr (-)}{}_i B^i_{\scr (-)}$. Second, invoke Urbantke's theorem \cite{Freidel, Urbantke} to parametrize the two triples of 2-forms as
\be\label{BbS}
B^i_{\scr (+)}(b,e) = \eta b^i_a \Sigma^a_{\scr (+)}(e), \qquad 
B^i_{\scr (-)}(\bar b,\bar e) = \bar\eta\bar b^i_a \Sigma^a_{\scr (-)}(\bar e).
\ee
In this expressions, $\eta$ and $\bar\eta$ are signs, the spacetime scalars $b$ and $\bar b$ form unimodular 3-by-3 real matrices, and
\be\label{Sigma}
\Sigma^i_{\scr (\pm)}(e) = e^0 \w e^i \pm \f12 \eps^i{}_{jk} \, e^j\w e^k
\ee
are Plebanski's 2-forms \cite{Plebanski, Capo2, Mike}. 
A straighforward calculation then checks that the metrics for these tetrads, $g(e)$ and $\bar g(\bar e)$, are precisely the two Urbantke metrics defined in \Ref{gpm}:
\be\label{gUpmeval}
g^{\scr (+)}_{\mu\nu} = g_{\mu\nu}, \qquad 
g^{\scr (-)}_{\mu\nu} = \eta \bar g_{\mu\nu}. 
\ee
Combining the two steps, one gets the generic parametrization
\be\label{ParamGen}
B^{IJ} =  P^{IJ}_{\scr (+)}{}_i \, b^i{}_a \Sigma_{\scr (+)}^a(e) + \eta P^{IJ}_{\scr (-)}{}_i \, \bar b^i{}_a \Sigma_{\scr (-)}^a(\bar e),
\ee
where we dropped an overall sign which turns out to be irrelevant: up to a global sign in the action, its effects can be absorbed in a redefinition of the parameters appearing in the potential.
The sets of scalars $b$ and $\bar b$, and the metrics $g(e)$ and $\bar g(\bar e)$, are all independent fields, parametrizing the 36 independent components of $B^{IJ}$. 
Notice that the decomposition \Ref{ParamGen} parametrizes $B^{IJ}$ in such a way that the right- and left-handed components 
$B^i_{\scr (\pm)}$ are also self- and antiself-dual, but with respect to two independent metrics $g_{\mu\nu}$ and $\bar g_{\mu\nu}$.

Using these metrics, one can solve the compatibility equation  $d_\om B=0$ and obtain a second order formulation of BF theory where the connection is on-shell.
The compatibility equation is reminiscent of Cartan's structure equation, $d_\om e=0$. If the tetrad $e$ is invertible, the equation admits a unique solution, given by the spin connection $\om(e)$. Similarly, $d_\om B=0$ admits a unique solution $\om(B)$, if $B$ satisfies certain non-degeneracy conditions.\footnote{Although non-degeneracy requirements breaks the purity of the initial configuration space, they do not change its dimensionality, nor spoil the shift symmetry. Hence, BF theory is topological also in the second order formalism.} 
One such condition is the invertibility of the two tetrads $e$ and $\bar e$.
Then, an explicit solution can be parametrized through the left-right splitting of the Lorentz algebra.
Accordingly, the compatibility equation splits into two independent $\mathfrak{su(2)}$ equations, $d_\om B^i=0$, whose solution is well-known \cite{Freidel,KrasnovEff,Deser,Halpern,Bengtsson},
\be\label{omB}
\om^i_\m(B) = \f1{4e} \eps^{\r\s\lambda\tau} B^{i}_{\lambda\tau} B_{j \r\m} \na^\n B^j_{\n\s}. 
\ee 
In each left/right sector independently, the indices are raised and lowered with the Urbantke metric, $e$ is the determinant of its tetrad, 
and $\na$ its covariant derivative.

Going back to the BF action, we can rewrite $P_\g=(1+1/\g)P_+ + (1-1/\g)P_-$, and use \Ref{omB} to write a second order action for BF theory,
\bea\label{Sl+r}
S_{BF}(B) = \f12\left(1+\f1\g\right) \int B^i_+\w F^i_+(\om_+(B_+))
- \f12\left(1-\f1\g\right) \int B^i_-\w F^i_-(\om_-(B_-)).
\eea
Using the parametrization \Ref{ParamGen}, the above second order BF theory is displayed as a bi-metric theory of gravity. The details of this lengthy calculation can be found in \cite{Speziale}. The result is best expressed in terms of the unimodular internal metrics 
\be
q_{ab} = b^i_a b^j_b \d_{ij}, \qquad \bar q_{ab} = \bar b^i_a \bar b^j_b \d_{ij},
\ee
and reads
\begin{subequations}\label{Sbi}\bea\label{Sbimetric}
S_{BF}(\eta,e^I_\mu, \bar e^I_\mu, q_{ab}, \bar q_{ab}) &=& 
\f14 \left(1+\f1\g\right) \int e R^{ab}_{\scr{(+)}}(e) (\hat q \d_{ab} - \hat q_{ab}) 
+ \f12 e \na^\mu q_{ab} C_{{\scr{(+)}}\ \mu\nu}^{abcd} \na^\nu q_{cd} \nn \\ &&  \hspace*{-1cm}
-\f\eta4 \left(1-\f1\g\right) \int \bar e R^{ab}_{\scr{(-)}}(\bar e) (\hat{\bar q} \d_{ab} - \hat{\bar q}_{ab}) 
+ \f12 \bar e \na^\mu \bar q_{ab} C_{{\scr{(-)}}\ \mu\nu}^{abcd} \na^\nu \bar q_{cd}.
\eea
Here $R^{ab}_{\scr{(\pm)}}(e)=(1/2) \Sigma^a_{\scr{(\pm)}}{}_{\mu\nu}(e)\Sigma^b_{\scr{(\pm)}}{}_{\rho\sigma}(e)R^{\mu\nu\rho\sigma}(e)
$ is the self-dual part of the Riemann tensor,
$q$ is the trace of $q_{ab}$, and $\hat q_{ab}$ its inverse. The kinetic term of the scalars is controlled by
$$
C^{abcd}_{{\scr{(\pm)}} \ \mu\nu}(e,q) \equiv \left(\d^{ad}\d^{bc} - \f12 \d^{ab}\d^{cd} \right) g_{\mu\nu} 
+ \left(\d^{bc} \eps^{ad}{}_g - \hat q^{bc} \eps^{adf} q_{fg} \right) \Sigma_{\scr{(\pm)}}^g{}_{\mu\nu}(e).
$$
Here and in the following, the internal indices $i$ or $a$ are contracted with the identity metrics.

The potential term can also be expressed in terms of the bi-metric parametrization. One finds
\begin{align}\nn
& Q_{ij}^2=4 e^2 ( q_{ab}^2-\f13q^2 ), && Q_{ij}^2=4 \bar e^2 ( \bar q_{ab}^2-\f13 \bar q^2 ),
\\\nn & Q^2=(2eq-2\bar e\bar q)^2, && Q_\star^2=(2eq+2\bar e\bar q)^2,
&& {\mathbbm Q}_{ij}^2 = 16 \ell_{ij}^2,
\end{align}
where 
\be\nn
\ell^{ij} \equiv \f18 b^i_a \bar b^j_b \, \eps^{\mu\nu\rho\sigma} \Si^a_{\mu\nu}(e) \bar \Si^b_{\rho\sigma}(\bar e).
\ee
Therefore, the potential \Ref{Vm} is given by
\bea\label{Vbimetric}
V(\eta,e^I_\mu, \bar e^I_\mu, q_{ab}, \bar q_{ab}) = -\Lambda (2eq+2\bar e\bar q) + 
\f{1 }{4(2eq+2\bar e\bar q)} \Big[ 4 a_1 e^2 q_{ab}^2 + 4 a_2 \bar e^2 \bar q_{ab}^2 \nn\\
- \f43\Big(a_1 - \f{a_4}2\Big) e^2 q^2 - \f43\Big(a_2 - \f{a_4}2\Big) \bar e^2 \bar q^2 + 8a_3 \ell_{ij}^2 - \f43 a_4 e\bar e q \bar q \Big].
\eea\end{subequations}
The two pieces \Ref{Sbi} give the bi-metric interpretation of the modified Plebanski action \Ref{S}.
The main difference with respect to the standard bi-metric theories is the presence of the additional scalars $q_{ab}$.\footnote{Another difference is that here both metrics have the same cosmological constant, but this can be easily amended, and independent constants can be obtained simply adding a term $\tl\Lambda Q$ to \Ref{Vm}.} 
However, a first indication that the extra scalars might be auxiliary fields comes from a counting of degrees of freedom: \Ref{Sbi} has in general eight degrees of freedom, precisely as bi-metric theories of gravity with generic potential \cite{Damour}. Furthermore, the expansion around a bi-flat background gives the same interpretation of the degrees of freedom, namely those of a massless and a massive spin-2 particle plus a scalar mode \cite{Speziale}. Upon such linearization, the scalar fields can be integrated out, and \Ref{Vm} gives a mass term for one of the two gravitons.\footnote{This situation is analogous to the auxiliary role of the scalars in the self-dual case \cite{KrasnovEff}. These facts suggest that it should be possible to dispose of the $b$ scalars, by solving their equations of motions, and identify the physical degrees of freedom within the metric tensors only. This can unfortunately only be done in perturbation theory, given the non-linearity of the field equations. For the self-dual case, it has been studied in \cite{KrasnovEff}}

\section{Bi-flat background and linearized BF theory}

It is immediate to see that the bi-flat configuration $b_a^i=\d_a^i$, $g_{\m\n}=\d_{\m\n}$,
and the same for the barred quantities, is a solution of the BF field equations.
We then define the perturbations
\be\label{pert}
b^i_a =\d^i_a + c^i_a, \qquad g_{\mu\nu} = \d_{\mu\nu} + h_{\mu\nu}
\ee
and the same for barred quantities.
The flat metric background can be used to introduce a tetrad $\d^I_\m$, and unambiguously project spacetime indices into internal indices. Using this, we can write the induced expansion on $B$ as
\be\label{BeC}
B^{IJ}_{\m\n} = \d_{\m K} \d_{\n L} \, \eps^{IJKL} + \d_{\m K} \d_{\n L} \, C^{IJKL},
\ee
where the perturbation $C$ has a certain dependence on $c, \bar c, h$ and $\bar h$ that we do not need for the moment,
and we chose $\eta=-1$. For $\eta=1$ the zeroth order is $\d_{\m K} \d_{\n L} \, \d^{IJKL}$.

\subsection{Linearized on-shell connection}

The compatible linearized connection can be obtained expanding \Ref{omB} to leading order.
However, it is more instructive to solve directly the linearized equation, as it brings in further intuition on the relation between SO(4) BF theory and general relativity. In fact, restricting attention to the bi-flat solution, in which a unique background metric is singled out, allows us to avoid having to go through the decomposition in self-dual and antiself-dual sectors.
To see this, let us first rewrite \Ref{dB} in components,
\be\label{dBlin1}
\eps^{\m\n\r\s} \left( \p_\m B^{IJ}_{\n\r} + \om_\m^{IK} B^{KJ}_{\n\r} - \om_\m^{JK} B^{KI}_{\n\r} \right) = 0.
\ee
Then, we make use of the expansion \Ref{BeC}, which introduces the flat background tetrad $\d^I_\m$,
and look for a perturbative solution for $\om$. At zeroth order, $\om$ trivially vanishes, thus we can directly look at the
first order term, linear in $C$ and the connection perturbation $\d\om$. It is convenient to use the background tetrad to project the equation on internal indices only, and introduce 
the notation $w^{M,IJ} = \d^{M\m} \, \d\om_\m^{IJ}$. Using simple identities for the contraction of epsilon and delta tensors,
the first order of the equation reduces to the simpler form
\be
w^{[I,J]S} - w_K{}^{K[I} \, \d^{J]S} = Y^{S,IJ},
\ee
where $Y^{S,IJ}=(1/4) \eps^{SMNR} \p_M C^{IJ}{}_{NR}$.
This linear equation can be easily inverted if we decompose both $w$ and $Y$ into their irreducible components
${\bf (3/2,1/2)}\oplus{\bf (1/2,3/2)}\oplus {\bf (1/2,1/2)}\oplus {\bf (1/2,1/2)}$. This can be done using the orthogonal projectors $\Pb, \Ps$ and $\Ph$ (see appendix \ref{App} for details),\footnote{The irreducible components $\wb$, $\ws$ and $\wh$ satisfy
\be\label{wprop}
\wb_{A,BC} \d^{AB} = \eps^{ABCD} \wb_{A,BC}  = 0, \qquad \ws_B = \d^{AC} w_{A,BC},  
\qquad \wh^{D} =  \f16 \eps^{ABCD} w_{A,BC}.
\ee}
\be\label{wdec}
w_{A,BC} = \Big(\Pb + \Ps+ \Ph\Big)^{I,JK}_{A,BC} \, w_{I,JK} = \wb_{A,BC} + \f23 \d_{A[C} \ws_{B]} + \eps_{ABCD} \wh^{D}.
\ee
Inserting it into \Ref{dBlin1}, we find (in matricial notation, with implicit indices)
\be
\Big(-\f12 \Pb + \Ps+ \Ph\Big) \, w = Y,
\ee
where we used $\wb^{[I,J]K} = -(1/2)\wb^{K,IJ}$, which follows from the symmetries of $\bar w$. Hence, 
\be\label{wsol}
w = (-2\Pb + \Ps+\Ph) Y = -2Y + 3(\Ps+\Ph) Y.
\ee

\subsection{Linearized BF action}

Integrating by parts in the BF action \Ref{SBF}, and using the solution $\om(B)$ to $d_\om B=0$, we get
\be
S(B) = P_\g^{ABKL} \int \om^{IK}(B)\w \om_{IJ}(B)\w B_{AB}.
\ee
Since $\om(B)$ vanishes at zeroth order, the quadratic approximation comes from $B^{AB}_{\m\n} = \d_{\m C} \d_{\n D} \, \eps^{ABCD} $
and $w^{M,IJ} = \d^{M\m} \, \d\om_\m^{IJ}$. After some simple algebra, we arrive at the second order linearized action
\be\label{SC}
S^{(2)} = -\f12 \int w_{IJK} \, M_{A,BC}^{I,JK} \, w^{ABC}, \qquad 
M_{A,BC}^{I,JK} = 4 \d^{[K}_{[B} P_\g{}^{J]I}_{C]A}. 
\ee
Explicitly, in terms of the irreducible components,
\be\label{H2}
S^{(2)} = -\f12 \int \wb^2 - \f43 \ws^2 -12 \wh^2 + \f8\g \ws \cdot \wh +\f1{2\g} \eps^{ABCD} \wb_{I,AB} \wb_{I,CD} .
\ee
We remark that this is precisely the term quadratic in the connection fluctuations of the linearized Holst action for GR; see e.g. \cite{Benedetti}.
The difference is that here $w$ is not an independent variable, but rather $w=w(C)$ through \Ref{wsol}.
The relation shows that the linearized kinetic term of a bi-metric theory comes from a single Einstein-like Lagrangian,\footnote{Indeed, the $\g$-independent parts of \Ref{H2} are like Einstein's $\Gamma\Gamma$ action for the Lorentz connection.} where the connection is on-shell a function of two metrics.

\subsection{Bi-metric parametrization}

At this point we use the explicit dependence of $C$ on the perturbations \Ref{pert}, to rewrite \Ref{SC} for the metric perturbations.
One can easily check that the first order perturbation is given by
\be\label{Bpert}
C^{IJKL} =  2 \Pp^{IJ}{}_i \, \Pp^{KL}{}_a \, c^{ia} +2\eta  \Pm^{IJ}{}_i \, \Pm^{KL}{}_a \, \bar c^{ia}
+ 2\Pp^{IJ[K}{}_M \, h^{L]M} +2\eta \Pm^{IJ[K}{}_M \, \bar h^{L]M}
\ee
Notice that it is not symmetric under $IJ\mapsto KL$, consistently with the fact that $B$ and its perturbations have 36 independent components. The decomposition permits to appreciate explicitly the way $c, \bar c, h$ and $\bar h$ capture the independent components of $C$: decomposing \Ref{Bpert} in its irreducibles, we find
\be\label{ciccio}
c^{(ia)} \in {\bf (2,0)},
\qquad c^{[ia]} \in {\bf (1,0)},
\qquad h^+_{\m\n} \in {\bf (1,1)}\oplus {\bf (0,0)}
\qquad h^-_{\m\n} \in {\bf (1,1)}\oplus {\bf (0,0)}
\ee

This parametrization can now be used to evaluate the on-shell connection \Ref{wsol}, and its curvature. 
To that end, we introduce the perturbations
\be
q_{ab}=\d_{ab}+\chi_{ab}, \qquad
q= 3+\f12 \chi_{ab}^2,
\ee
and idem for the barred quantities. The second order approximation for $q$ follows from $\det(q_{ab})=1$. Looking directly at the curvature, we have
\bea
\d F^{IJ}_{\m\n}(\om(C)) &=& 2 \, \p_{[\m} w_{\n]}^{IJ}(C) = \d_{\m\n}^{ KL} \, \d F^{IJKL}(C),
\\ \label{F1}
\d F^{IJKL} &=&  \Pp^{IJ}{}_a \, \Pp^{M[K}{}_b \, \p^{L]} \p_M \, \chi^{ab} 
- \Pm^{IJ}{}_a \, \Pm^{M[K}{}_b \, \p^{L]} \p_M \, \bar\chi^{ab} \\\nn
&& + \f12 \d R^{IJKL}(h+\bar h) + \f14 \eps^{IJ}_{MN} \, \d R^{MNKL}(h-\bar h),
\eea
where $\d R^{IJKL} =\p_\nu\p_{[\rho} h_{\sigma]\mu}-\p_\mu\p_{[\rho} h_{\sigma]\nu}$ is the first order expansion of the Riemann tensor.
Unlike $C$, this tensor does not have all its possible components: the ${\bf (1,0)}\oplus {\bf (0,1)}$ components vanish, which is due to the special symmetries of the background.

These formulas can be easily reduced to the case of general relativity:
In the presence of simplicity constraints, $\chi=\bar\chi=0$ and $h=\bar h$, then $C^{IJKL}=-\eps^{IJM[K} h^{L]M}$, \Ref{wsol} correctly gives the linearized spin connection $w_{I,JK}=\p_{[K}h_{J]I}$, where we used the symmetry of $h_{IJ}$, and finally $\d F^{IJKL}  =  \d R^{IJKL}(h)$.

Finally, the linearization of \Ref{Sbimetric} can be shown as in \cite{Speziale} to give
\be\label{Sbflin}
S_{BF}^{(2)}= \f12\left(1+\f1\g\right) \int {\cal L}^{(2)}_{\rm EH}\Big(h_{\mu\nu}+\chi_{\mu\nu}\Big)
- \f\eta 2\left(1-\f1\g\right) \int {\cal L}^{(2)}_{\rm EH}\Big(\bar h_{\mu\nu}+\bar\chi_{\mu\nu}\Big),
\ee
where 
\be
{\cal L}^{(2)}_{\rm EH}(h_{\m\n})\equiv -\f12 h_{\m\n} E^{\m\n\r\s} \square h_{\r\s}
=-\f 1 4 h_{\m\n,\lambda}^2+\f12 h_{\m\n,\n}^2-\f 1 2 h_{\m\n,\n}h_{,\n}+\f1 4 h_{,\m}^2
\ee
is the second order approximation of the Einstein-Hilbert action,
and
\be\label{chimunu}
\chi_{\mu\nu} \equiv  P^a_\eps{}_{\mu\rho} P^b_\eps{}_{\nu\sigma} \f{\p^\rho \p^\sigma}{\square} \chi_{ab}, \qquad P^a_\eps{}_{\mu\rho} = P^a_\eps{}_{IJ} \d^I_\mu\d^J_\rho,
\ee
is transverse, traceless and invariant under diffeomorphisms.

We see that the stability of the theory depends on both $\eta$ and the domain of $\gamma$. To have both kinetic terms in \Ref{Sbflin} positive, we can choose $\eta=-1$ and $|\gamma|>1$, or $0<\gamma<1$ and $\eta=1$. The remaining case, $-1<\gamma<0$ and $\eta=1$ leads to an overall minus sign and can be also treated if one reverses also the signs in the potential term.
Concerning the latter, the exact expression \Ref{Vbimetric} gives, applying standard formulas for the expansions (see \cite{Speziale} for details), 
\be\label{Vlin}
V^{(2)}= \f1{12} \left[a_1 \chi_{ab}^2 + a_2 \bar \chi_{ab}^2 
+ \f{3a_4-a_3}{8} (h-\bar h)^2 + \f{a_3}2 (h_{\m\n} - \bar h_{\m\n})^2 \right].
\ee

\section{Pauli-Fierz mass term}
\label{PFsec}

Let us consider non-singular potentials, that is exclude the cases where one or more of the parameters $a_i$ of \Ref{Vlin} vanish.
Since the potential term breaks the shift symmetry, we can now make the field redefinitions
\be
H_{\m\n} = h_{\m\n} + \chi_{\m\n}, \qquad \bar H_{\m\n} = \bar h_{\m\n} + \bar\chi_{\m\n}.
\ee
It is convenient to reabsorb the $\g$-prefactors of \Ref{Sbflin} in the fields in order to get the kinetic term in canonical normalization.
Further control can be obtained by diagonalizing the mass Lagrangian \Ref{Vlin}. It is easy to check that the diagonalization is obtained through
\be
H^{(+)}_{\m\n} = \f{|1+\g|}{\sqrt{2}|\g|} H_{\m\n} + \f{|1-\g|}{\sqrt{2}|\g|} \bar H_{\m\n}, 
\qquad H^{(-)}_{\m\n} = \f{\sqrt{|\g^2-1|}}{\sqrt{2}|\g|} (H_{\m\n} - \bar H_{\m\n}).
\ee
We also define
\be
\chi^{(\pm)}_{\m\n} = \f1{\sqrt{2}} \left(\chi_{\m\n} \pm \bar\chi_{\m\n}\right).
\ee  
Accordingly, the final form of the Lagrangian in the stable case reads
\bea\label{Lmass}
{\cal L} &=& \f12 \mathcal{L}^{(2)}_{\textrm {EH}}(H^{(+)}_{\m\n}) + \f12 \mathcal{L}^{(2)}_{\textrm {EH}}(H^{(-)}_{\m\n}) 
+\f1{12}\f{\g^2}{|\g^2-1|} \left[ \f{3a_4-a_3}4 H^{(-)}{}^2 + a_3 H^{(-)}_{\m\n}{}^2 \right] \label{LagChi}
\\\nn
&&  -\f{a_3}{6} \f{|\g|}{\sqrt{|\g^2-1|}} H^{(-)}_{\m\n} \chi_{(-)}^{\m\n}
+ \f{a_3}{12} \chi^{(-)}_{\m\n}{}^2 + \f{a_1+a_2}{24} (\chi^{(+)}_{\m\n}{}^2 + \chi^{(-)}_{\m\n}{^2})+\f{a_1-a_2}{12} \chi^{(+)}_{\m\n} \chi_{(-)}^{\m\n}.
\eea
This is a Lagrangian for a massless and a massive graviton, although it is written in a non-standard form, due to the presence of the extra scalars, which at this stage only enter algebraically. As we will see below, this does not affect the choice of parameters leading to the PF mass term, but only the final value of the mass, which can already at this point be seen to be $\g$-dependent. This $\g$-dependence is brought in by the need to canonically normalize and diagonalize the mass term.

We can now study the particle content of this lagrangian, and identify the choice of parameters in the potential which give the Pauli-Fierz mass term.
To that end, let us look at the field equations:  
taking variations of \Ref{Lmass} with respect to the $H^{(\pm)}$ and $\chi^{(\pm)}$ fields, we have
\begin{subequations}\label{linEq}\begin{align}\label{boxH+}
& {\mathcal E}^{\r\s}_{\m\n} H^{(+)}_{\r\s} = 2\kappa^2 \tau^{(+)}_{\m\n}, \\ \label{boxH-}
& {\mathcal E}^{\r\s}_{\m\n} H^{(-)}_{\r\s} 
-\f{1}{3} \f{2\g^2}{|\g^2-1|}\left[ a_3 H^{(-)}_{\m\n} +\f{3a_4-a_3}4 \d_{\m\n}H^{(-)}\right]
+\f{a_3}{3}\sqrt{\f{\g^2}{|\g^2-1|}}\chi^{(-)}_{\m\n}= 2\kappa^2 \tau^{(-)}_{\m\n}, \\ \label{boxchi+}
& \f 1 {12}(a_1+a_2)\chi^{(+)}_{\m\n} + \f 1 {12}(a_1-a_2)\chi^{(-)}_{\m\n} = \kappa^2 \sigma^{(+)}_{\m\n}, \\ \label{boxchi-}
& \f1{12}(2{a_3}+a_1+a_2)\chi^{(-)}_{\m\n} + \f1{12}(a_1-a_2)\chi^{(+)}_{\m\n}-\f{a_3}{6}\sqrt{\f{\g^2}{|\g^2-1|}} H^{(-)}_{\m\n}{}^{TT} = \kappa^2 \sigma^{(-)}_{\m\n}.
\end{align}\end{subequations}
Here ${\mathcal E}^{\r\s}_{\m\n} $ is a short-hand notation for the linearized Einstein equations, 
\be
{\mathcal E}^{\r\s}_{\m\n}h_{\r\s}=-\square h_{\m\n}+2\partial^\r\partial_{(\m}h_{\n)\r}-\partial_\m\partial_\n h-\d_{\m\n}\partial^\r\partial^\s h_{\r\s}+\square h\d_{\m\n}
\ee
and in \Ref{boxchi-},
$H^{(-)}_{\m\n}{}^{TT}$ 
is the spin-2, transverse-traceless component of $H^{(-)}_{\m\n}$, which appears due to the transverse-tracelessness of $\chi^{(-)}_{\m\n}$.
For completeness, and future reference, we have also included on the right-hand side some source terms of matter fields, with universal coupling $\kappa^2$, although a precise discussion of matter coupling goes beyond the scope of this paper. 
Let us just notice that the $\chi$'s affect the coupling to matter. In particular, we see that the case $a_1\neq a_2$, in which parity-breaking effects can be potentially expected, leads to a mixing between the $\scr{(+)}$ and $\scr{(-)}$ fields, and their matter sources.

The first equation is the standard linearized Einstein's equation for the massless graviton. The second is the equation for the massive graviton, with a generic mass term, and some of the scalars acting as sources. This is the equation we want to focus on.
The last two equations are algebraic and, provided the determinant is non-zero, i.e. $\Delta\equiv~2a_1a_2+a_3(a_1+a_2)\neq 0$, can be solved for $\chi^{(+)}$ and $\chi^{(-)}$.  Equation \Ref{boxH-} then yields a closed equation for $H^{(-)}$. The case $\Delta=0$ will be discussed below.
We thus get
\be\label{TTmas}
{\mathcal E}^{\r\s}_{\m\n} H^{(-)}_{\r\s} + a H^{(-)}_{\r\s} + b \d_{\m\n} H^{(-)} + c H^{(-)}_{\m\n}{}^{TT} = 
\kappa^2 \tilde \tau_{\m\n}, 
\ee
where
\be
a= -\f23\f{\g^2}{|\g^2-1|} a_3, \qquad b = -\f23\f{\g^2}{|\g^2-1|} \f{3a_4-a_3}4,
\qquad c= \f23\f{\g^2}{|\g^2-1|} \f{a_3^2 (a_1+a_2)}{\Delta},
\ee
and
\be
\tilde \tau= \tau^{(-)} -  \f{4|\g|}{\sqrt{|\g^2-1|}}\f{a_1+a_2}{\Delta}a_3 \s^{(-)}- \f{4|\g|}{\sqrt{|\g^2-1|}} \f{a_2-a_1}{\Delta}a_3 \s^{(+)}.
\ee

To study what quanta are propagated by this equation, we decompose $H^{(-)}_{\m\n}$ using the projectors $P^{(s)}$ in individual spin components,\footnote{The projectors are given in terms of
$\om\equiv \f{\p_\m \p_n}{\square}$ and the tranverse projector $D_{\m\n}=\d_{\m\n}-\om_{\m\n}$, as
\bea\nn
P^{(2)} = \f12 (D_{\m\r} D_{\n\s} + D_{\m\s} D_{\n\r} ) -\f13 D_{\m\n}D_{\r\s},  \qquad   P^{(0)}_s = \f13 D_{\m\n}D_{\r\s},  \\\nn
P^{(1)} = \f12 (D_{\m\r} \om_{\n\s} + D_{\m\s} \om_{\n\r} + D_{\n\r} \om_{\m\s} + D_{\n\s} \om_{\m\r}),  \qquad   P^{(0)}_l =  \om_{\m\n} \om_{\r\s},
\eea
and satisfy ${\mathbbm 1} = P^{(2)}+P^{(1)}+P^{(0)}_s+P^{(0)}_l$ in the space of symmetric tensors. 
The spin-0 space is two-dimensional, and one needs also the off-diagonal operators
\[
\qquad P^{(0)}_{sl} = \f1{\sqrt{3}} D_{\m\n} \om_{\r\s}, \qquad P^{(0)}_{ls} = \f1{\sqrt{3}} \om_{\m\n} D_{\r\s}.
\]
}
\be
H^{(-)}_{\m\n} = \left[ P^{(2)} + P^{(1)} + P^{(0)}_l + P^{(0)}_s \right] H^{(-)}_{\m\n},
\ee
where $H^{(-)}_{\m\n}{}^{TT} = P^{(2)} H^{(-)}_{\m\n}$ is the spin-2 transverse-traceless part already introduced, and the rest is a spin-1 and two spin-0  components.
A standard procedure \cite{Nieuw} then gives, after decoupling the scalar components,
\begin{subequations}\bea
&& \left[-\square +a+c\right]P^{(2)} H^{(-)}_{\m\n} = \kappa^2 P^{(2)}\tilde\tau_{\m\n} \\
&& a P^{(1)} H^{(-)}_{\r\s} = \kappa^2 P^{(1)}\tilde\tau_{\m\n} \\
&& \left[2(a+b)\square+a^2+4ab\right]P^0_s H^{(-)}_{\m\n} = 
\kappa^2 (a+b)P^{(0)}_s \tilde\tau_{\m\n}-\sqrt{3}bP^{(0)}_{sl} \tilde\tau_{\m\n}\\
&& \left[2(a+b)\square+a^2+4ab\right]P^0_l H^{(-)}_{\m\n} = \left[2\square + a + 3b\right]P^0_l\tilde\tau_{\m\n}-\sqrt{3}bP^0_{ls}\tilde\tau_{\m\n}.
\eea\end{subequations}
Therefore, the propagator in Fourier space is given by
\be
\Pi = \frac{P_2}{k^2+a+c}+\frac{P^1}{a}+\frac{(a+b)P^0_s-\sqrt{3}b(P^0_{ls}+P^0_{sl})+(-2k^2+a+3b)P^0_l}{-2(a+b)k^2+a^2+4ab}.
\ee
One can check that the residue at $k^2=-(a+c)$ is positive definite, but the residue, for $a+b\neq 0$, at $\frac{a^2+4ab}{2(a+b)}$, is negative. 
Propagation of the ghost mode is avoided for $a=-b$, the PF mass term. In terms of the parameters of the potential $V(B)$, this condition gives
\be\label{result}
a_4=-a_3,
\ee
with arbitrary $a_1$ and $a_2$. 
This is precisely the form that could have been guessed directly from \Ref{Lmass}, thus the scalar perturbations $\chi$'s do not affect this aspect of the theory. They do, however, affect the final value of the mass of the spin-2 particle, which for generic parameters ends up being
\be\label{mass}
M^2 = -\f43 \f{\g^2}{|\g^2-1|}\f{a_1 a_2 a_3}{\Delta}.
\ee
The parameters $a_1,a_2,a_3$ have to be chosen such that the graviton is not a tachyon, in order to keep the bi-flat solution stable. Notice furthermore that the Immirzi parameter enters with the same algebraic combination as in the effective coupling to fermions \cite{fermions}.\footnote{For $\g^2=1$, we are dealing with the modified self-dual Plebanski action, which only has 2 degrees of freedom \cite{Krasnov2}. In complete agreement, the mass diverges and the second graviton stops propagating.}

Let us now consider also the $\Delta=0$ case. 
Since we are excluding singular potentials with one or more coefficients vanishing, this leaves only configurations with
\be
\label{nospin2}
a_3=-\f{2a_1a_2}{a_1+a_2}, \qquad a_1+a_2\neq 0. 
\ee
For these values, the mass \Ref{mass} diverges and consequently $P^{(2)}H^{(-)}_{\mu\nu}$ stops propagating. 
The condition for the further non propagation of the scalar ghost is again \Ref{result}.\footnote{This is confirmed looking again at \Ref{linEq}: there, $\chi^{(\pm)}$ decouple from $H^{(-)}{}^{TT}$ and the result is that $H^{(-)}{}^{TT}$ is then determined from the sources $\sigma^{(\pm)}$, and does not propagate. Its trace part, the scalar ghost, is also eliminated for the same value \Ref{result}.} 
Hence, with \Ref{nospin2} and \Ref{result} holding, the linearized theory only propagates a spin-2 massless particle, $H^{(+)}$, and it is equivalent to linearized general relativity (provided the right coupling to matter is chosen). 


\section{Some simple singular potentials}\label{SecNP}

We have identified the choice \Ref{result} of parameters in the quadratic potential \Ref{Vm} that leads to a Pauli-Fierz mass term, featuring the absence of the ghost at the linearized level. 
We have furthermore identified the choice \Ref{nospin2} leading to the absence of the second spin-2 particle. 
These choices of coefficients affect the counting of degrees of freedom of the linearized theory.
We can now ask whether any of these reductions of degrees of freedom survives when interactions are included. To answer this question, we need to look at the canonical analysis of the exact theory. As recalled above, the non-perturbative number of degrees of freedom is determined by \Ref{gradL}. If there are no zero modes in the Hessian, then the interacting theory has eight DOFs, regardless of the linearized analysis. This is precisely the case:
when the coefficients $a_i^{-1}$ in \Ref{fAf} are non-zero, $\Lambda^{(2)}$ has maximal rank, independent of fine-tunings such as \Ref{result} or \Ref{nospin2}. 
Hence, even if the choice \Ref{result} eliminates the ghost at the linear level, it reappears through the interactions, precisely as the Boulware-Deser ghost of massive gravity. In other words, it is a necessary but not sufficient condition.
The same argument applies  to the absence of the massive spin-2 mode when $\Delta=0$: higher-order interactions will generate these additional degrees of freedom.  

In order to obtain fewer degrees of freedom at the full, non-perturbative level, one needs potentials with singular Hessian. The quadratic case \Ref{fAf} provides an interesting case study, as a singular Hessian can be obtained easily setting some of the coefficients $a_i^{-1}$ to zero. 

\subsection{Self-dual theory}
Set $\Lambda=\Lambda(\vphi^{ij})$, or $a_2^{-1}=a_3^{-1}=a_4^{-1}=0$ in the quadratic case. Only the ${\bf (2,0)}$ part of the simplicity constraints are relaxed. Consequently, the dynamical fields are a unique metric, with tetrad $e$, and a unique set of scalars $b$. The $B^{IJ}$ can be parametrized as 
$$
B^{IJ} = \left[ \Pp^{IJ}{}_i\Pp^a{}_{KL} b^i_a +\eta \Pm^{IJ}{}_{KL}  \right] e^K\w e^L.
$$
The on-shell right-handed connection is still given by \Ref{omB}, whereas the left-handed one is directly (the projection of) the spin connection $\om(e)$. Solving the algebraic field equations for $\vphi^{ij}$, we recover a generic potential $eV(m^{ij})$.
The resulting action $S(b^i_a,e^I_\mu)$ has the structure of the modified self-dual Plebanski theories studied by Krasnov \cite{Krasnov2}.  Hence, this class of potentials gives only 2 degrees of freedom.

Notice also that if we take $\gamma^2=1$ (or $-1$ in the Lorentzian case), then we also fall into the self-dual modified theory. In this case, no matter the form of the potential, there are only two degrees of freedom.

\subsection{Scalar-tensor theory}
Set $\Lambda=\Lambda(\d^{IJKL}\phi_{IJKL})$, or all but $a_4^{-1}$ vanishing in the quadratic case. The scalar simplicity constraint is the only one relaxed. The dynamical field are now a unique metric plus a scalar field, the conformal factor relating the right- and left-handed Urbantke metrics. 
The resulting theories have only 3 degrees of freedom, and the scalar mode is not necessarily a ghost: full stability can be obtained for suitable values of $\gamma$ and $\eta$. Such scalar-tensor sector is studied in detail in \cite{Beke}.

\subsection{Scalar constraint and unimodular massive gravity}

Set $\Lambda=\Lambda(\vphi^{ij},\bar\vphi^{ij},\psi^{ij})$, or $a_4^{-1}=0$ in the quadratic case. The scalar simpicity constraint is the only one present, and a single zero mode is introduced in the Hessian. A priori, this is a promising case because one could expect the reintroduced scalar constraint to kill one non-perturbative DOF.
However, this is not the case, as it is evident already in the perturbative expansion around the bi-flat background. Proceeding as above, one gets exactly the formula \Ref{Lmass}, the only difference being that now $H^{(-)}_{\m\n}$ is traceless. 
Projecting on the spin components, its propagator reads
\be\label{blurp}
\Pi = \frac{P_2}{k^2+a+c}+\frac{P^1}{a}+\frac{P^0_l+P^0_s}{-\f1 2k^2+a}.
\ee
We see that we have again the scalar ghost, and furthermore it is now always dynamical, independent of any parameter fine-tuning. That is, this subclass is completely sick.
The result can be understood also from the viewpoint of the canonical analysis: while taking $a_4^{-1}=0$ we do achieve a singular Hessian $\Lambda^{(2)}$, its zero mode is not in the six-dimensional space of Lagrange multipliers which give genuine constraints.\footnote{With the linearized result \Ref{blurp}, we have recovered that unimodular massive gravity contains six DOFs, with the massive graviton and the Boulware-Deser ghost. This is also apparent in the ADM formalism, where the lapse function N is no longer an independent variable (since $N^2\det(q)=1$). In the massless case, the diffeomorphism constraint is still present as a primary constraint and the Hamiltonian constraint, with cosmological constant as an integration constant, appears as the secondary constraint generated by it \cite{Unruh}. In the massive case however, the only way to obtain a constraint is by fine-tuning the mass term such that $N$ appears linearly, which is of no help in the unimodular case.}
This negative result shows the non-triviality of finding a class of theories with 7 degrees of freedom, a problem open in bigravity theories, and to which the Plebanski description unfortunately does not seem to bring much novelty.
In particular, we can identify a specific algebraic difficulty: the most natural single constraint to be added, the scalar simplicity constraint, gives a zero mode lying in the wrong part of the Hessian.

\subsection{Genuine bi-gravity}
Set $\Lambda=\Lambda(\psi^{ij},\delta^{IJKL}\phi_{IJKL})$, or $a_1^{-1}=a_2^{-1}=0$ in the quadratic case.
The $(\bf{2,0}) \oplus \bf{(0,2)}$ components of the simplicity constraints are present, imposing  $q_{ab}=\d_{ab}$ and $\bar{q}_{ab}=\d_{ab}$. Fur such potentials, the scalar fields are explicitly eliminated, and a genuine bi-metric theory of gravity is obtained. With the potential \Ref{fAf}, we have
\bea\label{Sbigenuine}
S 
 & = & \f12\int \left(1+\f1\g\right) eR(e) - \eta \left(1-\f1\g\right) \bar{e}R(\bar{e})  \\ \nn && \quad
		+ \f1{12(e+\bar{e})}\left[a_3 e^2 ({\cal K}^2 - {\cal K}^\m_\n {\cal K}^\n_\m) +6a_4(e^2+\bar{e}^2) -12(a_3+a_4)e\bar{e}\right],
\eea
where ${\cal K}^\m_\n \equiv g^{\m\r} \bar{g}_{\r\n}$, and ${\cal K} \equiv \d^\n_\m {\cal K}^\m_\n$.
A generic $\Lambda(\psi^{ij},\delta^{IJKL}\phi_{IJKL})$ will result in a generic interaction term $V({\cal K})$,
which corresponds to the most general action for bi-metric theories of gravity.\footnote{For more on bi-gravities and their applications, see e.g. \cite{Blas,Nesti,Banados,Milgrom}.}

Linearization of \Ref{Sbigenuine} around the bi-flat background leads to 
\be
\mathcal{L} = \f12 \mathcal{L}^{(2)}_{\textrm {EH}}(h^{(+)}_{\m\n}) + \f12 \mathcal{L}^{(2)}_{\textrm {EH}}(h^{(-)}_{\m\n}) + 
\f{1}{12} \f{\g^2}{|\g^2-1|} \left(a_3h^{(-)}_{\m\n}+\f{3a_4-a_3}{4}h^{(-)}{}^2\right).
\ee
The PF mass term is again obtained for $a_3=-a_4$, with
\be
M^2_{PF}=-\f23 \f{\g^2}{|\g^2-1|} a_3.
\ee
This corresponds to the theory
\be
\f12\int eR(e)+\bar{e}R(\bar{e}) + \f{a_3}{12(e+\bar{e})}\left[ e^2 ({\cal K}^2 - {\cal K}^\m_\n {\cal K}^\n_\m)-6(e^2+\bar{e}^2)\right].
\ee
However, this theory has already been examined in the literature, and the canonical analysis also yields eight degrees of freedom \cite{Damour}. Once again, the ghost reappears in the interactions.
Let us add a couple of further remarks.
\begin{itemize}
\item The presence of the ghost in the genuine bi-gravity case suggests that it will also be present in the cases with only one of the auxiliary fields constrained, namely $a_1^{-1}=0,a_2^{-1}\neq 0$ or $a_1^{-1}\neq0,a_2^{-1}= 0$. Nevertheless, for completeness one can note that the same relation $a_3=-a_4$ has to be imposed in order to obtain the PF mass term. The respective masses are then given by
\be
M^2_{a_1^{-1}=0}=-\f43\f{\g^2}{|\g^2-1|}\f{a_2a_3}{2a_2+a_3},\qquad M^2_{a_2^{-1}=0}=-\f43\f{\g^2}{|\g^2-1|}\f{a_1a_3}{2a_1+a_3}.
\ee
Taking $a_4^{-1}=0$ in any of the cases with $a_1^{-1}a_2^{-2}=0$ is of no help either: for exactly the same reasons as in the case $a_1^{-1}a_2^{-2}\neq0$, the massive graviton is unimodular, and the ghost is already propagating at the linear level.

\item The fact that the choice of parameters leading to a Pauli-Fierz mass term, $a_3=-a_4$, is unchanged in the presence of the additional $(b,\bar b)$ scalar degrees of freedom (see previous Section), is further evidence that these fields, although can not be immediately integrated out, do not play much of a dynamical role, but just encode some additional self-interactions of the two metrics.

\end{itemize}

\bigskip

\noindent While the above list does not exhaust all the possible singular cases for the quadratic potential, it contains all cases in which the $(\bf{1,1})$ components of the constraints are relaxed, such that the two Urbantke metrics are independent and one can expect a bi-gravity theory. This shows how non-trivial it is to find a class with seven non-perturbative DOFs, that is generalizing the Pauli-Fierz mass term to the interacting theory. One might hope that a special, but still polynomial form of the potential with the right zero mode in the Hessian exists. However, the genuine bi-gravity case suggests that this is not possible. Indeed, a theory of massive gravity which appears to be non-perturbatively ghost free has been introduced and studied in a series of papers  \cite{deRham, Hassan1}, and its casting within a bi-metric theory of gravity also considered in \cite{Hassan2}.
In this approach, the generic non-perturbative Boulware-Deser ghost is bypassed by tailoring the non-perturbative mass term to have a singular Hessian with respect to the lapse field and shift vector used in the ADM analysis. 
The upshot of these results is that one needs specific potentials of $\sqrt{{\cal K}^\m_\n}$: The presence of the square root is necessary to achieve the ghost-freeness. This somewhat cumbersome construction can be translated into an equivalent non-polynomial potential $V(Q)$ in terms of Plebanski variables, and consequently, suggests that also the full modified Plebanski theory would require some complicated and non-polynomial potential to eliminate the ghost non-perturbatively. Notice also that the canonical Hamiltonian analysis is made rather complicated by the form of the potential, which has led to some debate in the literature on the effective ghost-freeness \cite{HassanNew}. Even translated in the Plebanski formalism, checking the ghost-freeness through the zero modes of the Hessian appears just as complicated. Indeed, it would be a palatable feature of the Plebanski formalism to shed some light on such unusual structure of the potential, but our analysis in this direction has produced no interesting results so far.

\section{Lorentzian theory and reality conditions}\label{secLor}

In this final section, we discuss a rather different aspect of the theory, which concerns its description in the case of Lorentzian signature.
The analysis performed above extends immediately to SO(3,1), the relevant group for Lorentzian signature, but there is a caveat: because the left-right splitting 
$\mathfrak{sl(2,\C)=su(2)\oplus su(2)}$ now involves a complexification of the $\mathfrak{su(2)}$ algebras,
reality conditions must be supplemented to make the metrics \Ref{gpm} real. In fact, for signatures $\sigma=\pm$, \Ref{gpm} reads
\be\label{gpmLor}
\sqrt{-g^{(\pm)}} \, g^{(\pm)}_{\mu\nu} = 
\f1{6} \d_{IN}\left(\d_{JMKL} \pm  \f{\sqrt{\sigma}}2 \eps_{JMKL}\right) B^{IJ}_{\mu\a}\tl B^{KL \a\b}B^{MN}_{\b\nu}.
\ee
In the Lorentzian case, $\sigma=-1$ and $\sqrt{\sigma}=i$.
The surretitious presence of the factor $i$ in this formula is what introduces the need of reality conditions. 
These can be borrowed from the self-dual formalism \cite{Capo2}, as was done in \cite{Speziale}. The resulting metrics are real and automatically Lorentzian,
as a consequence of the Urbantke theorem \cite{Freidel,Urbantke}, but one needs to work with a \emph{complex} field $B$,
as inspection of \Ref{gpmLor} immediately reveals.
It is then desirable to have an alternative strategy which gives real Lorentzian metrics while preserving the reality of $B$.

Assuming $B$ real, one has to look for an alternative metric decomposition. A natural possibility is to work with the real and imaginary parts of \Ref{gpm}.
However, it turns out to be more convenient to work with a slightly different pair of metrics, say $g_1$ and $g_2$, defined as 
\be\label{defg12}
-g^{(\pm)} \, g^{(\pm)}{}^{\mu\nu} = \pm \tl{\tl g}_1^{\mu\nu} + \sqrt{\sigma} \tl{\tl g}_2^{\mu\nu},
\ee
where the double tilde is to keep track of the fact that the right-hand side is a density-two pseudotensor.
Using the inverse of \Ref{gpmLor}, it is not hard to show that
\be\label{g12}
\tilde{\tilde g}_1^{\mu\nu} = 
\f1{6} \d_{IN} \d_{JMKL}  \tilde B^{IJ \mu\a}B^{KL}_{\a\b} \tilde B^{MN \b\nu}, \qquad 
\tilde{\tilde g}_2^{\mu\nu} = 
\f1{12} \d_{IN} \eps_{JMKL} \tl B^{IJ \mu\a}B^{KL}_{\a\b}\tl B^{MN \b\nu}.
\ee
These metrics have been considered for instance in \cite{CapoImm}. For both signatures, they are real for real $B$, without any additional conditions.
Notice that  \Ref{g12} is an equality between density-\emph{two} pseudotensors.

The use of \Ref{g12} instead of the Urbantke metrics \Ref{gpmLor} gives an alternative approach to the physical interpretation of the modified Plebanski theory,
and it  would be interesting to develop it further, and see whether one gets the same bi-metric-plus-scalar structure, or substantial differences appear. Ultimately, which reality conditions ($B$ complex and \Ref{gpmLor} real, as above, or $B$ real and \Ref{g12} as the fundamental metrics) are physically relevant depends on the form of the potential term and the coupling to matter. In particular, we postpone a study of the signatures of \Ref{g12} and how to parametrize $B$ in terms of them.
To illustrate nonetheless the validity of these alternative metrics, in the rest of this section we show that they can be used to provide a new parametrization of the solution to the compatibility equation, which does not rely on the left/right splitting, and does not require any reality conditions.
Explicitly, we will show below that the SO(4) compatibility equation
\be\label{dom}
d_\om B^{IJ}=0
\ee
is solved by
\be\label{totti}
\om^{AB}_\m = -\f18 ( \Id \tl{\tl g}_{1} +\sigma\star  \tl{\tl g}_{2})^{-1}{}^{AB}_{XY}{}_{\m\n}\, [ \Id \otimes \Id + \sigma \star\otimes\star]^{XY}_{PQ}{}^{IJ}_{RS} 
\, \tl B_{IJ}^{ \n\a} B^{PQ}_{\a\b} \p_\r \tl B^{RS \r\b}.
\ee
The solution can be checked to transform correctly under changes of gauge and diffeomorphisms,\footnote{Notice that the partial derivative enters as a divergence of a density-one bi-vector, so what we have is just a 3-form. Indeed, the whole expression can be also written in terms of forms.} and to reduce to the sum of left- and right-handed solutions \Ref{omB}.

\subsection{Solving the compatibility condition}

We will need first to prove the following identity,
\be\label{giovanni}
\left(\Id \tl{\tl g}_1 + \sigma \star \tl{\tl g}_2\right)^{EF}_{XY}{}^{\m\n} 
= \f18 f_{IJKLMN} \, [ \Id \otimes \Id + \sigma \star\otimes\star]^{EF}_{PQ}{}^{MN}_{XY} \, \tl B^{IJ \mu\a}B^{PQ}_{\a\b}\tl B^{KL \b\nu}, 
\ee
where
\be
f_{IJKLMN} = 4 \eta_{[M [I} \eta_{J] N] KL}
\ee
are the structure constants of the Lorentz group. 
The starting point is a similar formula holding in the SU(2) case,
\be\label{App1}
\d^l_k \tl {\tl g}_\eps{}^{\m\n} = \f12 \eps_{ijk} \tl B_\eps^{i\m\a} \tl B_\eps^{j\n\b} B_\eps^l{}_{\a\b},
\ee
where $\eps=\pm$ stands for self- or antiself-dual, and $\tl {\tl g}_{\eps}{}^{\m\n} = -g_\eps g_{\eps}{}^{\m\n}$ is the doubly densitized (inverse) Urbantke metric,
and $B^i_{\eps} = P_\eps{}^i_{IJ} B^{IJ}$.
The formula is easily proved using \Ref{BbS} and its (anti)self-duality.

Next, we observe that the SU(2) structure constants $\eps_{ijk}$ are the self-dual part of the Lorentz algebra ones $f_{IJKLMN}$, more precisely
\be
\eps_{ijk} = -\f{\eps}{2\sqrt{\sigma}} f_{IJKLMN} P_\eps^{IJ}{}_i P_\eps^{KL}{}_j P_\eps^{MN}{}_k.
\ee
Using this formula in \Ref{App1}, we get
\bea\label{App2}
\d^l_k \tl {\tl g}_\eps{}^{\m\n} &=& \f\eps2 f_{IJKLMN} P_\eps^{IJ}{}_{AB} \tl B^{AB\m\a} P_\eps^{KL}{}_{CD} \tl B^{CD\n\b} 
P_\eps^{MN}{}_{k} P_\eps{}_{PQ}^{l} B^{PQ}{}_{\a\b}  \nn\\
&=&  \f\eps2 f_{IJKLMN} \tl B^{IJ\m\a} \tl B^{KL\n\b} 
P_\eps^{MN}{}_{k} P_\eps{}_{PQ}^{l} B^{PQ}{}_{\a\b}, 
\eea
where in the first equality we used trivial properties of the projectors, and in the second the fact that SD/ASD is an orthogonal decomposition for the structure constants, thus a single projector on a pair of indices suffices.
The last formula can be equivalently rewritten as 
\be
2 P_\eps{}^{EF}_{GH} \tl {\tl g}_\eps{}^{\m\n} = 
 \f\eps2 f_{IJKLMN} P_\eps^{MN}{}_{GH} P_\eps{}_{PQ}^{EF} \tl B^{IJ\m\a} \tl B^{KL\n\b} B^{PQ}{}_{\a\b}.
\ee
Expanding out the projectors, and using the definition \Ref{g12}, we get
\bea
&& \left[\sqrt{\sigma} \left(\Id \tl{\tl g}_2 + \star \tl{\tl g}_1\right) + \eps \left( \Id \tl{\tl g}_1 + \sigma \star \tl{\tl g}_2 \right) \right]^{EF}_{GH}{}^{\m\n}  \nn\\\nn
&& \qquad = \f\eps8 f_{IJKLMN} \left[ \Id\otimes\Id +\sigma \star\otimes\star +\eps\sqrt{\sigma}\left( \Id\otimes\star+\star\otimes\Id \right) \right]^{MN\,EF}_{GH\,PQ} 
\, \tl B^{IJ \mu\a}B^{PQ}_{\a\b}\tl B^{KL \b\nu}.
\eea
Finally, subtracting the equation for $\eps=-1$ to the one for $\eps=1$, we arrive at \Ref{giovanni}.

We can now easily write down the solution of \Ref{dom}.
In components, the equation can be written as
\be
\p_\r \tl B_{IJ}^{\r\s} = f_{IJKLMN} \om^{MN}_\r \tl B^{KL \r\s}.
\ee
Then, we contract both sides of this expression with 
$[ \Id \otimes \Id + \sigma \star\otimes\star]^{XY}_{PQ}{}^{IJ}_{RS}  \tl B^{IJ \m\a} B^{PQ}_{\a\b}$,
and use the fact that $\star$ commutes with $f$ to get
\bea
&& [ \Id \otimes \Id + \sigma \star\otimes\star]^{XY}_{PQ}{}^{IJ}_{RS} \,  \tl B^{IJ \m\a} B^{PQ}_{\a\b} \p_\r \tl B_{IJ}^{\r\s} \nn\\\nn
&& \qquad = f_{IJKLMN}\,  \om^{MN}_\n \tl B^{KL \n\s}
\, [ \Id \otimes \Id + \sigma \star\otimes\star]^{XY}_{PQ}{}^{IJ}_{RS} \,  \tl B^{IJ \m\a} B^{PQ}_{\a\b} \\ \nn
&& \qquad =  -\f18\om^{AB}_\m ( \Id \tl{\tl g}_{1} +\sigma\star  \tl{\tl g}_{2})_{AB}^{XY}{}^{\m\n},
\eea
where in the last step we used \Ref{giovanni}. Inverting the last equation, we find the solution \Ref{totti}. Finally, let us add that a similar approach to solve the compatibility
equation for general SO(N) groups has been studied in \cite{Cuesta}.

\section{Conclusions}
In this paper, we have studied the modified non-chiral Plebanski action, generalizing the results of \cite{Speziale} in various directions.
First, we considered the most general quadratic potential in which all simplicity constraints are relaxed. In particular, we showed how to cure the presence of a linear ghost by identifying the values of the parameter leading to the Pauli-Fierz mass term. This is given by \Ref{result}.
Second, we included the Immirzi parameter in the analysis, and found that the mass of the second spin-2 particle depends on it, via the combination $\gamma^2/(\gamma^2-1)$. 
This might appear as a surprising result at first, since the reader might be familiar with the classical irrelevance of the Immirzi parameter in pure gravity. However, the parameter allows to differentiate the left and right sectors of the Lorentz algebra, and in the modified theory the two sectors carry independent degrees of freedom. The result displays the importance of the Immirzi parameter in the modified Plebanski theory. 
We also found values leading to the absence of the second spin-2 particle. These are given by \Ref{nospin2}.
Third, we provided many more details on the linearized analysis, including the structure of irreducible representation, the solution of the linearized spin connection, a simple relation between the second order linearized BF theory and general relativity, and the helicity decomposition of the field equations, needed to study the scalar mode.

Beyond the linearized approximation, we discussed how one can try to remove the ghost also non-perturbatively. While the Pauli-Fierz mass term is a necessary condition, it is not sufficient and a Boulware-Deser ghost can reapper through the interactions. We showed that this indeed happens for our quadratic potential with Pauli-Fierz mass term. Although it has been argued that a theory of massive gravity with a non-linear ghost can be physically viable, e.g.  \cite{Damour, Arkani, Mukhanov}, it is important to understand whether a non-perturbatively ghost-free theory exists. We argued that the most intuitive possibility, which is to reintroduce the scalar simplicity constraint, does not work, because of the structure of secondary constraints of the action. Then we showed how to link the modified Plebanski action to genuine bi-metric theories of gravity. For these, a non-perturbatively ghost-free construction has been recently presented in the literature \cite{deRham,Hassan1}, and it would be interesting to pursue further studies in this direction, and see whether the two approaches can shed light on each other. On the other hand, a non-perturbatively ghost-free subclass is obtained relaxing only the scalar simplicity constraint, and one obtains scalar-tensor theories \cite{Beke}.

Finally, we also presented an alternative solution to the compatibility condition, given in \Ref{totti}. The new solution does not rely on the left/right splitting of the algebra, but rather on a metric structure induced by the structure constants of the Lorentz group, leading to the key identity \Ref{giovanni}. The advantage of the new solution is that in the Lorentzian case it is automatically real for real $B$, whereas the standard solution needs a complex $B$ field and additional reality conditions to make the Urbantke metrics real.  This new construction could bring new insights in the interpretation of the modified non-chiral Plebanski action.

\bigskip

{\bf Acknowledgements}.
We are pleased to thank N. Van den Bergh for stimulating discussions, and Merced Montesinos for showing us a draft of his
paper \cite{Cuesta}. David is supported by the Research Foundation-Flanders (FWO). Simone is partially supported by the ANR Programme Blanc grant $LQG-09$. David and Giovanni wish to thank the CPT for hospitality during part of this work.

\newpage

\appendix
\section{Appendix}\label{App}

In this appendix, we collect our conventions and the explicit expressions for the various projectors used in the main body of the paper.

\subsection{Conventions}
\begin{itemize}

\item {Levi-Civita pseudo-tensor:}

\begin{align}\nn
&\eps^{0123}=1, &&
e = \f1{4!} \eps_{IJKL}\eps^{\m\n\r\s} e_\m^I e_\n^J e^\r_K e^\s_L,
\\\nonumber
& \eps_{\mu\nu\rho\sigma} = g_{\mu\a} g_{\nu\beta} g_{\rho\g} g_{\sigma\d} \eps^{\a\b\g\d},
&& \eps^{\mu\nu\rho\sigma} \eps_{\mu\nu\rho\sigma} = 4! g,
\\\nonumber
& \eps^{\m\n\r\s}e_\m^I e_\n^J = e\, \eps^{IJKL} e^\r_K e^\s_L, 
&& \f{1}{4e^2} \eps_{\m\n\a\b}\eps^{\r\s\a\b} = \d_{\m\n}^{\r\s}. 
\end{align}

\item {Self-dual projectors and their properties:}
\be\nn
P_{\eps}^{IJ}{}_{KL} = \f12\left(\d^{IJ}_{KL} + \f\eps2 \eps^{IJ}_{KL}\right),
\qquad
P_{\eps}^{IJ}{}_i = 2 P_{\eps}^{IJ}{}_{0i} = \d^{IJ}_{0i} + \f\eps2 \eps^{IJ}_{0i}
\ee
\begin{align}\nn
& \d^{ij} P_{ \eps}^{IJ}{}_i P_{ \eps}^{KL}{}_j = P_{ \eps}^{IJKL},
\qquad \d_{IJKL} P_{ \eps}^{IJ}{}_i P_{ \eps}^{KL}{}_j = \d_{ij},
\qquad \f12 \eps_{IJKL} P_{ \eps}^{IJ}{}_i P_{ \eps}^{KL}{}_j = \eps\d_{ij},
\\\nn
& P^i_{\eps}{}_{IJ} P^j_{\eps}{}^{J}{}_L = -\f14 \d^{ij} \d_{IL} -\f\eps2 \eps^{ijk} P^k_\eps{}_{IL}, \\\nn
& P^i_{\eps}{}_{IJ} P^j_{-\eps}{}^{J}{}_L = \f14 \d^{ij} \d_{IL} -\f12\d^{ij}_{(IL)}
+\d^0_{(I} P^i_\eps{}_{JL)} \d^{jJ} = \f14 \d^{ij} \Big(\d_{IL} - 2\d^0_I\d^0_L\Big) -\f12\d^{ij}_{(IL)}
+\f\eps2 \d^0_{(I} \eps^{0ij}{}_{L)}.
\end{align}

\item Plebanski 2-forms and their properties:
\begin{align}\nn
& \Si^i_\eps = P^i_{\eps}{}_{IJ} e^I\w e^J, \qquad 
\f1{2e}\eps^{\m\n}_{\r\s} \Si^i_\eps{}_{\m\n} = \eps \Si^i_\eps{}_{\r\s}, \\\nn
& \Si^i_\eps{}_{\mu\nu}(e) =2 P^i_\eps{}_{\mu\nu} 
= 2 e_{[\mu}^0 e_{\nu]}^i + \eps \eps^i{}_{jk} e^j_\mu e^k_\nu, \\\nn
&\Sigma^i_\eps(e) \w \Sigma_\eps^j(e) = \eps \, 2 e \d^{ij} \, d^4x, \\\nn &
\f12 \d_{ij} \Si^i_\eps{}_{\mu\nu} \Si^j_\eps{}_{\rho\sigma} = 
g_{\mu[\rho} g_{\sigma]\nu}+\f\eps{2e}\eps_{\mu\nu\rho\sigma},
\\\nn & 
\Si^i_\eps{}_{\mu\nu} \Si^j_{\eps}{}_{\rho\sigma} g^{\nu\sigma} = \d^{ij} g_{\mu\rho} + \eps \eps^{ij}{}_l \Si^l_\eps{}_{\mu\rho},
\\\nn & 
\Si^i_\eps{}_{\mu\nu} \Si^j_\eps{}^{\mu\rho} \Si^k_\eps{}_{\rho\sigma} =
\d^{ij} \Si^k_\eps{}_{\nu\sigma} - \d^{ik} \Si^j_\eps{}_{\nu\sigma} + \d^{jk} \Si^i_\eps{}_{\nu\sigma}
-\eps \eps^{ijk} g_{\nu\sigma}.
\label{triple}
\end{align}

\end{itemize}

\subsection{Connection components}
The connection can be decomposed in irreducible representations of SO(4)
\be\label{omreps}
\bf{(3/2,1/2)\oplus(1/2,3/2)\oplus(1/2,1/2)\oplus(1/2,1/2)}.
\ee
The decomposition into (parity-even) irreducibles is realized by the three orthogonal projectors
\be\label{Preps}
\bar{P}_{A,BC}^{I,JK} =  \d_A^I \d_{BC}^{JK} - \check{P}_{A,BC}^{I,JK}  - \hat{P}_{A,BC}^{I,JK},
\quad \check{P}_{A,BC}^{I,JK} =  \f23 \d_{A[C} \d_{B]}^{[J} \d^{K]I},
\quad \hat{P}_{A,BC}^{I,JK} =  \f16 \eps_{ABCD}\eps^{IJKD}.
\ee

\subsection{Curvature components}

A general tensor $t^{IJKL}$ has 36 components, transforming under the following irreps,
$$
{\bf (2,0)}\oplus{\bf (0,2)}\oplus{\bf (1,0)}\oplus{\bf (0,1)}\oplus{\bf (1,1)}\oplus{\bf (1,1)}
\oplus{\bf (0,0)}\oplus{\bf (0,0)}.
$$
The SO(4) projectors are:
\bea
\Pi_{\bf(2,0)} &=&  \Pp^{(IJ|}{}_{AB} \, \Pp^{|KL)}{}_{CD} - \f13 \Pp^{IJKL} \, \Pp{}_{ABCD} \\
\Pi_{\bf(0,2)} &=&  \Pm^{(IJ|}{}_{AB} \, \Pm^{|KL)}{}_{CD} - \f13 \Pm^{IJKL} \, \Pm{}_{ABCD} \\
\Pi_{\bf(1,0)} &=&  \Pp^{[IJ|}{}_{AB} \, \Pp^{|KL]}{}_{CD} \\
\Pi_{\bf(0,1)} &=&  \Pm^{[IJ|}{}_{AB} \, \Pm^{|KL]}{}_{CD} \\
\Pi_{\bf(1,1)} &=&  \Pp^{(IJ|}{}_{AB} \, \Pm^{|KL)}{}_{CD} + \Pp^{(IJ|}{}_{CD} \, \Pm^{|KL)}{}_{AB} \\
\Pi_{(\overline{\bf 1,1})} &=& \Pp^{[IJ|}{}_{AB} \, \Pm^{|KL]}{}_{CD} - \Pp^{[IJ|}{}_{CD} \, \Pm^{|KL]}{}_{AB} \\
\Pi_{\bf(0,0)} &=&  \f1{6} \d^{IJKL} \d_{ABCD} \\
\Pi_{(\overline{\bf 0,0})} &=&  \f1{4!} \eps^{IJKL} \eps_{ABCD} \\
\eea
Using the fact that self-dual indices are in correspondence with SU(2) indices, we can write the composition in terms of the SU(2) indices as
\bea
t^{IJKL} &=& \sum_{i} (\Pi_i \, t)^{IJKL} = 
\Big(\Pp^{IJ}{}_i \Pp^{KL}{}{}_j - \f13 \d_{ij} \Pp^{IJ}{}_k \, \Pp^{KL}{}_k \Big) \, t^{(ij)} \\\nn 
&& + \Big(\Pm^{IJ}{}_i \Pm^{KL}{}{}_j - \f13 \d_{ij} \Pm^{IJ}{}_k \, \Pm^{KL}{}_k \Big) \, \bar t^{(ij)}  
+ \Pp^{[IJ|}{}_i \Pp^{|KL]}{}{}_j \, t^{[ij]} + \Pm^{[IJ|}{}_i \Pm^{|KL]}{}{}_j \, t^{[ij]}  \\\nn 
&& + \Pp^{(IJ|}{}_i \Pm^{|KL)}{}_j \, {\mathbbm t}^{ij} + \Pp^{[IJ|}{}_i \Pm^{|KL]}{}_j \, \bar{\mathbbm t}^{ij} 
+\f16 \d^{IJKL} \, t + \f1{12} \eps^{IJKL} \, t_\star,
\eea
where we introduced the following irreducible components:
\bea
{\bf(2,0)} &\ni& t_{(ij)} = \Big( \Pp^{(IJ|}{}_i \, \Pp^{|KL)}{}_j - \f13 \d_{ij} \Pp^{IJ}{}_k \, \Pp^{KL}{}_k \Big) \, t_{IJKL} \\
{\bf(0,2)} &\ni& \bar t_{(ij)} =\Big( \Pm^{(IJ|}{}_i \, \Pm^{|KL)}{}_j - \f13 \d_{ij} \Pm^{IJ}{}_k \, \Pm^{KL}{}_k \Big) \, t_{IJKL} \\
{\bf(1,0)} &\ni& t_{[ij]} = \Big( \Pp^{[IJ|}{}_i \, \Pp^{|KL]}{}_j \Big) \, t_{IJKL}\\
{\bf(0,1)} &\ni& \bar t_{[ij]} = \Big( \Pm^{[IJ|}{}_i \, \Pm^{|KL]}{}_j \Big) \, t_{IJKL} \\
{\bf(1,1)} &\ni&  \mathbbm t_{ij} = 2 \Big( \Pp^{(IJ|}{}_i \, \Pm^{|KL)}{}_j \Big) \, t_{IJKL} \\
({\overline{\bf 1,1}}) &\ni& \bar{\mathbbm t}_{ij} = 2 \Big( \Pp^{[IJ|}{}_i \, \Pm^{|KL]}{}_j \Big) \, t_{IJKL} \\
{\bf(0,0)} &\ni& t = \d^{IJKL} \, t_{IJKL} \\
(\overline{\bf 0,0}) &\ni& t_\star = \f12 \eps^{IJKL} \, t_{IJKL} \\
\eea

From these we compute the ten general quadratic invariants,
\begin{align}
& Q_1(t) = t_{IJKL} \, t^{IJKL} 
= t_{(ij)}^2 + t_{[ij]}^2 + \bar t_{(ij)}^2 + \bar t_{[ij]}^2 + 
\f12 {\mathbbm t}_{ij}^2 +\f12 \bar{\mathbbm t}_{ij}^2 +\f16 t^2 +\f16 t_\star^2, \\
& Q_2(t) = \f12 \eps^{KL}_{MN} \, t_{IJKL} \, t^{IJMN} 
= t_{(ij)}^2 + t_{[ij]}^2 - \bar t_{(ij)}^2 - \bar t_{[ij]}^2 
- {\mathbbm t}_{ij}  \bar{\mathbbm t}^{ij} + \f13 t  t_\star, \\
& Q_3(t) = t_{IK} \, t^{IK} = \f14 t^2 + \f12 t_{[ij]}^2 + \f12 \bar t_{[ij]}^2 +\f14 {\mathbbm t}_{ij}^2, \\
& Q_4(t) = t^2, \\
& Q_5(t) = t_\star^2, \\ 
& Q_6(t) = t t_\star, \\
& Q_7(t) = t_{IJKL} \, t^{KLIJ} = 
t_{(ij)}^2 - t_{[ij]}^2 + \bar t_{(ij)}^2 - \bar t_{[ij]}^2 + 
\f12 {\mathbbm t}_{ij}^2 -\f12 \bar{\mathbbm t}_{ij}^2 +\f16 t^2 +\f16 t_\star^2, \\
& Q_8(t) = t_{IK} \, t^{KI} = \f14 t^2 - \f12 t_{[ij]}^2 - \f12 \bar t_{[ij]}^2 +\f14 {\mathbbm t}_{ij}^2, \\
& Q_9(t) = \f12 \eps^{IJ}_{MN} \, t_{IJKL} \, t^{MNKL} = 
t_{(ij)}^2 + t_{[ij]}^2 - \bar t_{(ij)}^2 - \bar t_{[ij]}^2 
+ {\mathbbm t}_{ij}  \bar{\mathbbm t}^{ij} + \f13 t  t_\star, \\
& Q_{10}(t) = \f12 \eps^{IJ}_{KL} \, t_{IJMN} \, t^{MNKL} =
t_{(ij)}^2 - t_{[ij]}^2 - \bar t_{(ij)}^2 + \bar t_{[ij]}^2 + \f13 t  t_\star,
\end{align}
In the symmetric case, $\phi^{IJKL}=\phi^{KLIJ}$, there are only six independent invariants, since 
$Q_7=Q_1$, $Q_{10}=Q_9=Q_2$, $Q_8=Q_3$. If furthermore $\phi_\star=0$ is imposed, only the four invariants from \Ref{Qinv} remain.


\end{document}